# Multisensor Management Algorithm for Airborne Sensors Using Frank-Wolfe Method

**Youngjoo Kim\* and Hyochoong Bang**

Department of Aerospace Engineering, Korea Advanced Institute of Science and Technology
**\*Corresponding author: yjkim@ascl.kaist.ac.kr**

**Abstract:** This study proposes an airborne multisensor management algorithm for target tracking, taking each of multiple unmanned aircraft as a sensor. The purpose of the algorithm is to determine the configuration of the sensor deployment and to guide the mobile sensors to track moving targets in an optimal way. The cost function as a performance metric is defined as a combination of the D-optimality criterion of the Fisher information matrix. The convexity of the cost function is proved and the optimal solution for deployment and guidance problem is derived by the Frank-Wolfe method, also known as the conditional gradient descent method. An intuitive optimal approach to deal with the problem is to direct the sensor to the optimal position obtained by solving a nonlinear optimization problem. On the other hand, the proposed method takes the conditional gradient of the cost function as the command to the deployed sensors, so that the sensors are guaranteed to be in the feasible points and they achieve the current best performance. Simulation results demonstrate that the proposed algorithm provides better performance than directing each sensor to its optimal position.

## 1. Introduction

Multisensor system has emerged as an important research subject in military as well as civilian applications. Whereas a single sensor generally can obtain limited information, multiple sensors are able to provide sufficient information obtained from different viewpoints and focuses. Given the advancement in sensor technologies, the amount of data the sensors can process has increased. This motivates researches on automatic management of multiple sensors to improve overall perception performance. Multisensor management is formally described as a system or a process that attempts to manage a set of sensors in a dynamic, uncertain environment, to improve the performance of data fusion [1]. Details of purpose, role, architecture, and problem of the multisensor management are well described in [1-3].

Some multisensor management algorithms have been developed within the domain of antisubmarine warfare [4, 5]. In particular, a general methodology for the management of multisensor systems can be found in [4]. Expanding the single-target tracking scheme, a problem of managing an array of sensors in order to track multiple targets was considered in [6, 7]. These works provide convincing answers to this application area in regard to efficient and autonomous deployment of sonobuoy resources in submarine tracking. Important steps of the multisensor management technique are to quantify and control the accuracy of the target state estimation. The Fisher information matrix (FIM) provides the means of achieving this aim in an efficient way [8, 9]. The inverse of the FIM denotes a lower bound on the performance of an estimator.

The purpose of this study is to develop a multisensor management algorithm applicable to airborne warfare, considering each unmanned aircraft as a sensor system. Our previous work [10] has attempted to achieve the goal, but it is limited to the management of stationary sensors. One major difference of the unmanned aircraft from the sonobuoys is that the former are mobile sensors. In order to obtain better performance, the sensors can move to appropriate positions to chase the moving targets. The fact calls our attention to the problem in the sensor management as to determining the appropriate positions of the deployed sensors. We have addressed path planning of a mobile sensor where the location of sensor deployment is different of that of an operational area [11]. Another recent work [12] has addressed control of multiple mobile sensors using a potential field obtained by gradient of a goal function, but it is limited to maneuvers in 2D space without consideration of sensor coverage. An airborne sensor needs to be in the feasible region where the measurement is available.

By using FIM to quantify the performance, in this paper, we propose a multisensor management algorithm for determining 1) when a new sensor is required, 2) where to deploy the new sensor, and 3) how to guide the deployed sensors. The first two decisions, of when and where to deploy a new sensor, have been addressed in [10]. In this study, on the other hand, we imposed constraints to the optimization problems. That is, we took sensor coverage in account for making decisions of deployment location and guidance of the sensors. Our contribution lies in the method to provide the guidance command to the deployed sensors. We formulated the problem of finding the best sensor location as a constrained convex optimization problem and provided the negative conditional gradient of the cost function as the guidance command, which is the solution of the subproblem of Frank-Wolfe optimization technique. Solving the subproblem every time step significantly reduced computation time while the sensors approached the optimal location, maintaining the targets in the sensors' coverage. The proposed multisensor management algorithm was verified by numerical simulations with stationary and moving targets.

The remainder of the paper is structured as follows. Section 2 presents the problem formulation for which the multisensor management algorithm is designed. Section 3 addresses how to define the performance metric and its convexity is proved therein. In Section 4, the multisensor management algorithm is given. The algorithm consists of determining the configuration of the sensor deployment and generating command to the deployed sensors. Section 5, some examples are given to demonstrate the effectiveness of the proposed algorithm. Summary and conclusions are presented in Section 6.

**2. Problem Formulation**

In this section, we present the problem formulation for the management algorithm for an airborne multisensor system to track multiple ground targets. It is assumed that each ground target is moving independently, and the number of the targets is $L$. The number of the unmanned aircraft is $N_{max}$ and that of the deployed is $N$. Suppose that each unmanned aircraft is equipped with a sensor system, which provides angle and range measurements. For example, one can use a radar system, a LIDAR system, or a camera system with terrain elevation data for the framework to be described in this paper. The purpose of the multisensor management is to obtain the best estimate of the target position. Each sensor system in this paper is assumed able to detect the false measurement and distinguish multiple measurements from the multiple targets. The modelling methodology for the false alarms and measurement association, generally considered in radar systems, can be referred to [4, 6, 7].

A major characteristic of the airborne multisensor system is that the sensors are mobile. In this study, we restrict the unmanned aircraft as a rotary-wing type that is able to hover and move to every direction unlike fixed-wing aircraft. This helps us to focus on the multisensor management, not the path planning. A fusion center as a sensor manager, which can be a ground control station or an unmanned aircraft, collects sensory data and determines when, where, and for which target group the unmanned aircraft should be deployed. At the same time, each mobile sensor is directed to chase the moving targets after its deployment by continuously calculating appropriate directions to track the targets to achieve better performance. Within the problem domain described above, there are three inter-related problems to be addressed.

The first problem is the time for the new sensors to be deployed, given the current estimate of the target state. The target tracking algorithm is supposed to satisfy a criterion as the lower bound of the state estimate performance. The time will be determined as the time when the accuracy of the state estimate is lower than a prespecified criterion.

The second problem is to determine the configuration of the next deployment; the target group to be tracked and the location of the newly deployed sensor. In other words, given a certain time for the next deployment, and the predicted target state at that time, the algorithm determines which targets should be tracked by the new sensor and where the sensor should be positioned in an optimal way. An optimization problem for maximizing information of targets the new sensor gathers should be solved to determine the configuration of the next deployment.

The third problem is the guidance of each sensor to track the moving targets. That is, the direction in which each sensor should move is supposed to be determined by solving the optimization problem.

*2.1. Target Model*

In this paper, the target system is described as a stochastic vector state space model. Each target is assumed to be moving independently with a constant velocity. The system model for each target $l = 1, 2, \ldots, L$ can be described as a near-constant-velocity model [13] by

$$\boldsymbol{X}_{k+1}^l = F_k^l \boldsymbol{X}_k^l + G_k^l \boldsymbol{v}_k^l \tag{1}$$

The target state at time $k$ is defined as $\boldsymbol{X}_k^l = [x_k, y_k, z_k, \dot{x}_k, \dot{y}_k, \dot{z}_k]^T = [\boldsymbol{p}_k, \boldsymbol{v}_k]^T$. The process noise $\boldsymbol{v}_k^l$ obeys a zero-mean Gaussian distribution whose covariance is expressed as

$$Q_k^l = \begin{bmatrix} (\sigma_x^l)^2 & 0 & 0 \\ 0 & (\sigma_y^l)^2 & 0 \\ 0 & 0 & (\sigma_z^l)^2 \end{bmatrix} \tag{2}$$

where $[\sigma_x^l \ \sigma_y^l \ \sigma_z^l]^T$ denotes the standard deviation of the noise and $\Delta t$ is the sampling interval. The matrices $F_k^l$ and $G_k^l$ are constructed as

$$F_k^l = \begin{bmatrix} I_{3\times3} & I_{3\times3}\Delta t \\ 0_{3\times3} & I_{3\times3} \end{bmatrix}$$
$$G_k^l = \begin{bmatrix} I_{3\times3} \frac{\Delta t^2}{2} \\ I_{3\times3}\Delta t \end{bmatrix} \tag{3}$$

and $I_{m\times n}$ is an $m$ by $n$ identity matrix. The covariance of $G_k^l \boldsymbol{v}_k^l$ is $G_k^l Q_k^l (G_k^l)^T$.

*2.2. Observation Model*

Consider the observation model regarding the measurement from the target $l$ by the sensor $n = 1, 2, \ldots, N$ given by

$$\boldsymbol{Z}_k^{n,l} = \boldsymbol{h}_k^{n,l}(\boldsymbol{X}_k^l) + \boldsymbol{\omega}_k^{n,l} \tag{4}$$

Let $\mathbb{T}_k^n$ be a set of targets assigned to the sensor $n$ at time $k$. Each sensor is supposed to track the target $l \in \mathbb{T}_k^n$ and provides measurement to the fusion center. Invalid measurements and measurements from other targets are discarded.

Each sensor provides angle measurements in azimuth and elevation, and a range measurement, as $\boldsymbol{Z}_k^{n,l} = [Az, El, R]^T$. Each measurement from the sensor $n$ tracking the target $l$ is expressed as

$$\boldsymbol{Z}_k^{n,l} = \begin{bmatrix} Az \\ El \\ R \end{bmatrix} + \boldsymbol{\omega}_k^{n,l} = \begin{bmatrix} \tan^{-1}\left(\frac{x^l - x^n}{y^l - y^n}\right) \\ \tan^{-1}\left(\frac{z^l - z^n}{\sqrt{(x^l - x^n)^2 + (y^l - y^n)^2}}\right) \\ \sqrt{(x^l - x^n)^2 + (y^l - y^n)^2 + (z^l - z^n)^2} \end{bmatrix} + \boldsymbol{\omega}_k^{n,l} \tag{5}$$

where $\omega_k^{n,l}$ is the zero-mean Gaussian measurement noise whose covariance matrix is $R_k^{n,l}$. The corresponding observation matrix $H_k^{n,l}$ is obtained by

$$H_k^{n,l} = \frac{\partial h_k^{n,l}(X_k^l)}{\partial X_k^l} \tag{6}$$

Note that $H_k^{n,l} = \partial h_k^{n,l}/\partial X_k^l$ has zero elements in the last three columns as

$$H_k^{n,l} = \begin{bmatrix} \frac{\partial h_k^{n,l}(X_k^l)}{\partial p_k^l} & \frac{\partial h_k^{n,l}(X_k^l)}{\partial v_k^l} \end{bmatrix} = \begin{bmatrix} \frac{\partial h_k^{n,l}(X_k^l)}{\partial p_k^l} & 0_{3\times 3} \end{bmatrix} \tag{7}$$

Letting $x = x^l - x^n$, $y = y^l - y^n$, $z = z^l - z^n$, the observation matrix can be expressed as

$$H_k^{n,l} = \begin{bmatrix} \frac{y}{x^2+y^2} & \frac{-x}{x^2+y^2} & 0 & \\ \frac{-xz}{(x^2+y^2+z^2)\sqrt{x^2+y^2}} & \frac{-yz}{(x^2+y^2+z^2)\sqrt{x^2+y^2}} & \frac{1}{\sqrt{x^2+y^2}} & 0_{3\times 3} \\ \frac{x}{\sqrt{x^2+y^2+z^2}} & \frac{y}{\sqrt{x^2+y^2+z^2}} & \frac{z}{\sqrt{x^2+y^2+z^2}} & \end{bmatrix} \tag{8}$$

The measurement is available only when the target $l$ is in the coverage of the sensor. For simplifying the problem, assume each sensor always looks downward. The sensor position is restricted as

$$\begin{aligned} \left|\tan^{-1}\left(\frac{x^l-x^n}{z^l-z^n}\right)\right| &< \frac{\theta_x}{2} \\ \left|\tan^{-1}\left(\frac{y^l-y^n}{z^l-z^n}\right)\right| &< \frac{\theta_y}{2} \\ z^n &> H_{min} \end{aligned} \tag{9}$$

where $\theta_x$ and $\theta_y$ denote angle of views of the sensor in x and y direction, respectively. $H_{min}$ denotes the minimum altitude of the sensor. By rearranging (9), a set of linear constraints for the sensor position is obtained as

$$\begin{aligned} \pm(x^l - x^n) - (z^l - z^n)\tan\frac{\theta_x}{2} &< 0, \\ \pm(y^l - y^n) - (z^l - z^n)\tan\frac{\theta_y}{2} &< 0, \\ z^n > H_{min} \quad \text{for all } l &\in \mathbb{T}_k^n \end{aligned} \tag{10}$$

The above equations will be used as the constraints of the optimization problems in section 4.

*2.3. Sequential Extended Kalman Filter*

In this paper, we adopted the sequential extended Kalman filter (EKF) to estimate the target state since every measurement and state transition are independent and the observation matrix can be expressed as the first-order derivative of the observation model. Brief introduction to the sequential EKF is given here and the details can be referred to the literature [14]. Its time update step is the same as that of typical EKF. The measurement update step of the sequential EKF is as follows.

$$\begin{aligned} \widehat{X}_{k+1,i}^+ &= \widehat{X}_{k+1,i-1}^+ + K_{k+1,i}\left(Z_{k+1,i} - h_i(\hat{X}_{k+1,i-1}^+)\right) \\ K_{k+1,i} &= P_{k+1,i-1}^+ H_{k+1,i}^T \left(H_{k+1,i} P_{k+1,i-1}^+ H_{k+1,i}^T + R_{k+1,i}\right)^{-1} \\ P_{k+1,i}^+ &= P_{k+1,i-1}^+ - K_{k+1,i} H_{k+1,i} P_{k+1,i-1}^+ \end{aligned} \tag{11}$$

where $\widehat{X}_{k+1,0}^+ = \widehat{X}_{k+1}^-$ and $\widehat{X}_{k+1}^+ = X_{k+1,I}^+$, and so does for $P$. In the above, $i = 1,2,\ldots,I$ denotes available measurement at the time step.

## 3. Performance Metric

*3.1. Fisher Information Matrix*

After the target state is estimated, the fusion center calculates the FIM for the multisensor system. Let $X_k$ be an unknown, random state vector to be estimated. Its unbiased estimate is defined as $\hat{X}_k(Z_k)$ when the measurement $Z_k$ is obtained. Then, the posterior Cramer-Rao lower bound (PCRLB) is defined to be the inverse of the FIM $J_k$. It provides a lower bound of the error covariance matrix of $\hat{X}_k(Z_k)$ as

$$C_k \triangleq \mathbb{E}\left\{[\hat{X}_k(Z_k) - X_k][\hat{X}_k(Z_k) - X_k]^T\right\} \geq J_k^{-1} \tag{12}$$

where $\mathbb{E}$ denotes expectation over $X_k$ and $Z_k$. The inequality in the above equation means that $C_k - J_k^{-1}$ is a positive semi-definite matrix.

The posterior FIM $J_{k+1}$ can be evaluated by the following formula given in [8, 9] as

$$J_{k+1} = (Q_k + F_k J_k^{-1} F_k^T)^{-1} + J_{Z,k+1} \tag{13}$$

The first term of the FIM $J_{k+1}$ contains the prior information of the target states at time $k+1$. The initial FIM is defined as $J_0 = (P_0)^{-1}$. It is assumed that the sensors can identify each target so that the measurement-to-target associations are known. If the association is unknown, its effect to the FIM can be evaluated by an information reduction matrix [6].

Note that the matrix $J_{Z,k+1}$ in (13) gives the measurement contribution to the FIM. For calculating the measurement contribution to the FIM, consider the observation model regarding the measurement of the target $l$ by the $n$th sensor given by (5) revisited as

$$Z_k^{n,l} = h_k^{n,l}(X_k^l) + \omega_k^{n,l} \tag{14}$$

where $h_k^{n,l}$ is a nonlinear function relating the target state and the measurement. $\omega_k^{n,l}$ is a zero-mean Gaussian measurement noise vector whose covariance is $R_k^{n,l}$. By expanding the discussion about simplifying the measurement contribution given in [4], $J_{Z,k+1}^{n,l}$ can be expressed as

$$J_{Z,k+1}^{n,l} = \left(H_{k+1}^{n,l}\right)^T \left(R_{k+1}^{n,l}\right)^{-1} \left(H_{k+1}^{n,l}\right) \tag{15}$$

Since the sensors have independent measurement processes, the measurement contribution $J_{Z,k+1}^l$ can be written as

$$J_{Z,k+1}^l = \sum_{n=1}^N J_{Z,k+1}^{n,l} \tag{16}$$
$$\text{for } l \in \mathbb{T}_{k+1}^n \text{ and } w_{k+1}^l \neq 0$$

where $J_{Z,k+1}^l$ denotes the measurement contribution to the FIM of tracking the target $l$. We adopted a weight vector $w$ to contain tracking priority on each target. For example, $w_k = [2,1,1,0]$ if an operator attempts to track the first three targets with higher priority on the first target. If a sensor is tracking the fourth target, the information of the fourth target will be ignored.

It is important to make the sensor management algorithm predictive because it needs sufficient time to plan and make the deployment. Therefore, if we perform the optimization at time $k$, it is necessary to predict the FIM in the subsequent sampling times $k+1, \dots$ for determining the configuration and the time of the next deployment. Given the FIM $J_k$ at time $k$, we can determine the predictive FIM $J_{k+r}$ at each time $k+r$, $r = 1,2,\dots$ by using the Riccati-like recursion, (13).

*3.2. Performance Score*

In this section, two kinds of scores are introduced. One is a measure of the amount of information of a certain target the fusion sensor gathers, called 'information score'. The other, called 'performance score', is a measure of the performance of a certain sensor to track the targets assigned to the sensor. Note that the optimization problem in this paper is formulated as a minimization problem, so that the lower score implies better performance.

3.2.1. Information of Each Target

Given the FIM, denoted as $J_{k+r}^l$, for tracking the target l at time $k + r$, the information score for the target position can be defined by using the D-optimality criterion as

$$c_{k+r}^l = -\ln \det(\Psi_{k+r}^l) \tag{17}$$

where $\Psi_{k+r}^l$ is a FIM for target position estimation taken as the first $3 \times 3$ elements in the FIM $J_{k+r}^l$ as

$$J_{k+r}^l = \begin{bmatrix} \Psi_{k+r}^l & J_{12} \\ J_{21} & J_{22} \end{bmatrix} \tag{18}$$

and $J_{12}$, $J_{21}$, $J_{22}$ are $3 \times 3$ matrices. The D-optimality criterion is based on the determinant of the FIM, which corresponds to the volume of the uncertainty ellipsoid. In addition to this, common criteria for optimal observation can be found in various literature [15, 16]. The individual information score $c_{k+r}^l$ is used in determining the time for the new deployment.

3.2.2. Performance of Each Sensor

In target state estimation, a general goal is to minimize the target position estimation error. In tracking multiple targets, we are interested in maximizing overall performance of each sensor for tracking targets assigned to it. The performance score will be used in determining both configurations of the new deployment and the guidance command to track the moving targets in an optimal way. The cost function describing the tracking performance is defined based on the measurement contribution to the FIM $J_{Z,k+1}^{n,l}$ where the state for the FIM is taken as the sensor state $X_{k+1}^n$ and the observation matrix is calculated as $H_{k+1}^{n,l} = \partial h_{k+1}^{n,l} / \partial X_{k+1}^n$.

The tracking performance score of the $n$th sensor at time $k + 1$ is defined as

$$s_{Z,k+1}^n = \sum_{l=1}^L c_{Z,k+1}^{n,l} w_{k+1}^l \quad \text{for } l \in \mathbb{T}_{k+1}^n \tag{19}$$

where

$$c_{Z,k+1}^{n,l} = -\ln \det(\Psi_{Z,k+1}^{n,l}) \tag{20}$$

and $\Psi_{Z,k+1}^{n,l}$ denotes the measurement contribution to the FIM of the senor n estimating the position of the target l at time $k + 1$. Note that, because the measurements have no observability in velocity, $H_{k+1}^{n,l} = \partial h_{k+1}^{n,l} / \partial X_{k+1}^n$ has zero columns as (7) and the measurement contribution to the FIM has a form of

$$J_{Z,k+1}^{n,l} = (H_{k+1}^{n,l})^T (R_{k+1}^{n,l})^{-1} (H_{k+1}^{n,l}) = \begin{bmatrix} \Psi_{Z,k+1}^{n,l} & 0_{3\times3} \\ 0_{3\times3} & 0_{3\times3} \end{bmatrix} \tag{21}$$

The performance score of the sensor $n$, (19), is a function of the sensor position and formulated as a weighted sum of the performance scores of the sensor $n$ to track the target $l$, denoted as $c_{Z,k+1}^{n,l}$. The information of the target whose weight is 0 is discarded. Only the measurement contribution term of (13) is used in defining the cost function because the information affected by the system model is independent of the sensor configuration. Note that as the determinant of the FIM becomes larger, the performance scores, (17) and (19), become lower. Thus, the global minimum of (19) is the optimal position of the sensor.

3.2.3. Convexity of the Performance Score

Before solving the optimization problem of target tracking, the convexity of the cost function has to be supported by theory. This subsection is dedicated to prove that the performance score of target tracking, $s_{Z,k+1}^n$, is convex. The definitions and theorems introduced in the following paragraphs are adopted from the literature [17-20] and some notations are adjusted according to the conventions of this paper.

**Definition 1 (Convex set [18]).** Let $\mathbb{S}$ be a subset of $\mathbb{R}^n$. We say that $\mathbb{S}$ is a convex set if
$$\alpha \boldsymbol{x} + (1-\alpha)\boldsymbol{y} \in \mathbb{S}, \; for \; all \; \boldsymbol{x}, \boldsymbol{y} \in \mathbb{S} \; and \; \alpha \in [0,1].$$

**Definition 2 (Convex function).** A function $f : \mathbb{S} \mapsto \mathbb{R}$ is called convex over a convex set $\mathbb{S}$ if
$$f(\alpha \boldsymbol{x} + (1-\alpha)\boldsymbol{y}) \leq \alpha f(\boldsymbol{x}) + (1-\alpha)f(\boldsymbol{y}), \; for \; all \; \boldsymbol{x}, \boldsymbol{y} \in \mathbb{S} \; and \; \alpha \in [0,1].$$

The cost function $s_{Z,k+1}^n$ defining the performance of target tracking has its domain, $\mathbb{S}$, as a set of positions of the sensor n, i.e., $\boldsymbol{p}_{k+1}^n \in \mathbb{S} = \mathbb{R}^3$. The set $\mathbb{S}$ is a convex set, and we need to prove the function $s_{Z,k+1}^n : \mathbb{S} \mapsto \mathbb{R}$ is convex over $\mathbb{S}$.

Firstly, let us deal with the function $c_{Z,k+1}^{n,l} : \mathbb{S} \mapsto \mathbb{R}$ defined in (20). It has been proved that the D-optimality criterion is convex in general as described in the following theorem.

**Theorem 1 (Convexity of the D-optimality criterion [20]).** The function $\Lambda : \Psi \mapsto -\ln\det(\Psi)$ is convex if $\Psi$ is positive semi-definite and strictly convex if $\Psi$ is positive definite.

Therefore, it indicates that the function $c_{Z,k+1}^{n,l}$ is convex if the FIM of the observation by the sensor $n$ tracking the target $l$, denoted as $\Psi_{Z,k+1}^{n,l}$, is positive semi-definite. The definition of positive definiteness is introduced below.

**Definition 3 (Positive definite [17]).** A matrix $\Psi$ is positive definite if $\boldsymbol{a}^T \Psi \boldsymbol{a} > 0$, and positive semi-definite if $\boldsymbol{a}^T \Psi \boldsymbol{a} \geq 0$ for all $\boldsymbol{a} \neq \boldsymbol{0}$.

Define

$$H_{k+1}^{n,l} = \begin{bmatrix} \frac{\partial h_{k+1}^{n,l}(\boldsymbol{p}_{k+1}^n)}{\partial \boldsymbol{p}_{k+1}^n} & 0 \end{bmatrix} = \begin{bmatrix} \boldsymbol{h}_1^T & 0_{1\times 3} \\ \vdots & \vdots \\ \boldsymbol{h}_m^T & 0_{1\times 3} \end{bmatrix} \quad (22)$$

where $\boldsymbol{h}_i$ are $3 \times 1$ vectors where $i = 1, \ldots, m$ and $m$ is the number of measurements of the sensor $n$. Consequently, the measurement contribution to the target position can be expressed as

$$\Psi_{Z,k+1}^{n,l} = \sum_{i=1}^m \sigma_i^{-2} \boldsymbol{h}_i^T \boldsymbol{h}_i \quad (23)$$

Choose any $3 \times 1$ vector $\boldsymbol{a} \in \mathbb{R}^3, \boldsymbol{a} \neq \boldsymbol{0}$, and multiply both sides of $\Psi_{Z,k+1}^{n,l}$ by $\boldsymbol{a}$ as

$$\boldsymbol{a}^T \Psi_{Z,k+1}^{n,l} \boldsymbol{a} = \sum_{i=1}^m \sigma_i^{-2}(\boldsymbol{a}^T \boldsymbol{h}_i)^2 \geq 0 \quad (24)$$

Therefore, $\Psi_{Z,k+1}^{n,l}$ is positive semi-definite, and we can conclude $c_{Z,k+1}^{n,l}$ is convex.

By expending the above discussion, it can be shown that the Fisher information matrix is positive semi-definite in general. If $\Psi_{Z,k+1}^{n,l}$ is full-rank, besides, $\text{null}\left(\left(\Psi_{Z,k+1}^{n,l}\right)^T\right) = \emptyset$. Thus, there does not exist a $\boldsymbol{a} \in \mathbb{R}^3, \boldsymbol{a} \neq \boldsymbol{0}$ such that $\boldsymbol{a}^T \boldsymbol{h}_i = 0$. Then, $\sigma_i^{-2}(\boldsymbol{a}^T \boldsymbol{h}_i)^2 > 0$ and $\Psi_{Z,k+1}^{n,l}$ is positive definite.

**Lemma 1 (Operation that preserves convexity [19]).** If $f_1, \ldots, f_m : \mathbb{S} \mapsto \mathbb{R}$ are convex, then the following function is convex as well.
$$X \mapsto \alpha_1 f_1(X) + \cdots + \alpha_m f_m(X) \quad \text{for } \alpha_1, \ldots, \alpha_m \geq 0$$

The performance score of the sensor $n$, $s_{Z,k+1}^n$, is calculated by a convex combination of $c_{Z,k+1}^{n,l}$ as shown in (19). Therefore, $s_{Z,k+1}^n$ is convex as well. It is concluded that there is a local minimum of $s_{Z,k+1}^n$ is also a global minimum.

## 4. Multisensor Management Algorithm

The proposed algorithm is described as a block diagram in Fig. 1. A cycle of the loop is executed every time a new measurement is arrived. The first and second steps, estimating target states and calculating the individual information score, are described in section 2 and 3 in detail. In this section, how to determine the configuration of the new sensor and how to guide the deployed sensors will be addressed. Given estimated target states, if any target exists where the information score of the target is larger than a prespecified value, i.e., $c_{k+r}^l > S_1$, the fusion center prepare a new sensor deployment as follows.

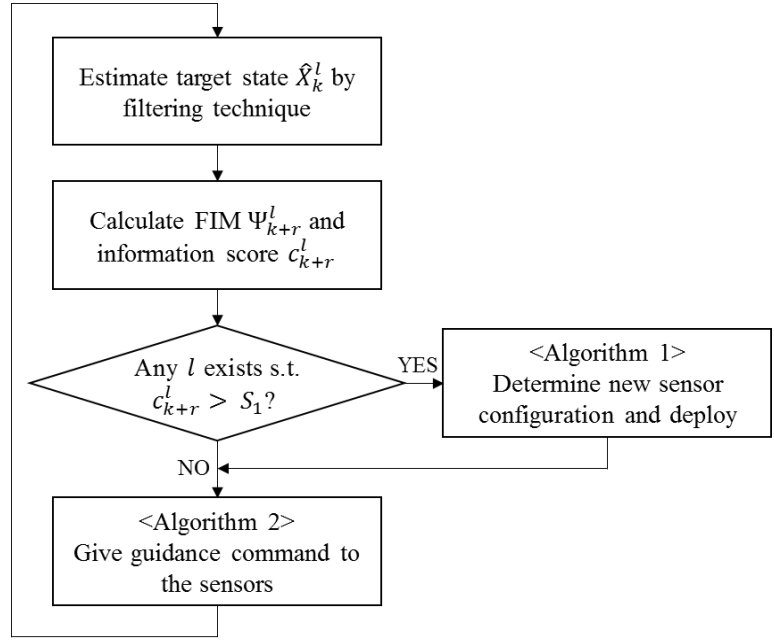

**Figure 1.** Block diagram of the proposed multisensor management algorithm.

### 4.1. Determining the Configuration of New Deployment

The configuration of the deployment means the location of the new sensor and the set of targets to be tracked by the sensor. For $L$ targets, there are possible $2^L - 1$ target groups. The number of the candidates can be reduced by selecting the candidate groups which contains the target $l^-$ whose information has been considered insufficient, i.e., $c_{k+r}^{l^-} > S_1$. Also, the target groups whose size is larger than a prespecified value $V$ are discarded. The maximum number of targets, $V$, reflects the limitation in processing power of sensor systems. Therefore, for each candidate of the target group $\mathbb{T}_{k+r}^{N+1,i} \in \{\mathbb{T}_{k+r}^{N+1,i} \mid l^- \in \mathbb{T}_{k+r}^{N+1,i} \text{ and } |\mathbb{T}_{k+r}^{N+1,i}| \leq V\}$ where $i = 1, \ldots, 2^L - 1$, the best score $s_{Z,k+r}^{N+1,i}$ and the corresponding sensor position are determined by solving an optimization problem. The candidate target group $\mathbb{T}_{k+r}^{N+1,i}$ is identified as the target group to be tracked if the best score of the configuration has the largest value among all the candidates. The algorithm for determining the new configuration is as follows.

**Algorithm 1 (Configuration of new deployment).**
*Get a set of candidate target groups:*
$\{\mathbb{T}_{k+r}^{N+1,i} \mid l^- \in \mathbb{T}_{k+r}^{N+1,i} \text{ and } |\mathbb{T}_{k+r}^{N+1,i}| \leq V\}$
*Set the best score* $s^* = \infty$
<u>*For*</u> *each* $\mathbb{T}_{k+r}^{N+1,i}$,
    Solve **Problem 1**:
        Get the best score $\left(s_{Z,k+r}^{N+1,i}\right)^*$,
        and the sensor position $\left(p_{k+r}^{N+1,i}\right)^* = \arg\left(s_{Z,k+r}^{N+1,i}\right)^*$
    $s^* = \min\left\{s^*, \left(s_{Z,k+r}^{N+1,i}\right)^*\right\}$, $p^* = \arg s^*$
*Deploy a new sensor at* $p^*$
$N = N + 1$
**END**

The convex optimization problem is defined as follows.

**Problem 1 (Best sensor position).**
*minimize* $s_{Z,k+r}^{N+1,i}\left(p_{k+r}^{N+1,i}\right)$
*subject to* $p_{k+r}^{N+1,i} \in P_{k+r}^{N+1,i}$

$P_{k+r}^{N+1,i}$ is defined as a set of sensor positions which satisfy the constraints specified in (10).

We solve Problem 1 using the conditional gradient method [18], also known as Frank-Wolfe method. The formulation for solving the optimization problem regarding a cost function $s$ which is a function of a position $p$ is given as follows. In order to generate a feasible direction $\bar{p}^j - p^j$ that satisfies the descent condition $\nabla s(p^j) \cdot (\bar{p}^j - p^j) < 0$, we use the method

$$p^{j+1} = p^j + \alpha^j(\bar{p}^j - p^j) \tag{25}$$

which is to solve the optimization problem

$$\begin{aligned} &\text{minimize } \nabla s(p^j) \cdot (p - p^j) \\ &\text{subject to } p \in P \end{aligned} \tag{26}$$

where $\alpha^j \in (0,1]$ is the stepsize which can be determined by some rules or set as a constant. Obtain $\bar{p}^j$ as the solution, that is,

$$\bar{p}^j = \arg\min_{p \in P} \nabla s(p^j) \cdot (p - p^j) \tag{27}$$

For Problem 1, $p^j$ in the above equations denotes sensor position $p_{k+r}^{N+1,i}$ at the $j$th iteration of the optimization, and $s(p)$ denotes the cost function of tracking performance. $\bar{p}$ is a point of $P$ that lies furthest along the negative gradient direction $-\nabla s(p)$. Because the subproblem, (26), is to minimize the linear approximation of the problem and $P$ is specified by linear constraints, it is a linear program, which is relatively easy to solve.

A corollary to Jacobi's formula [21] states that for any invertible matrix $\Psi$,

$$\frac{\partial}{\partial t} \ln \det \Psi(t) = \text{tr}\left(\Psi^{-1} \frac{\partial}{\partial t} \Psi\right) \tag{28}$$

where $\text{tr}(\cdot)$ denotes the trace of a matrix. Thus, the gradient of $c^{n,l} = -\ln \det \Psi^{n,l}$ can be expressed as

$$\nabla c^{n,l} = \left[\frac{\partial c^{n,l}}{\partial p^n}\right] = -\left[\text{tr}\left(\Psi^{-1} \frac{\partial}{\partial x} \Psi\right) \quad \text{tr}\left(\Psi^{-1} \frac{\partial}{\partial y} \Psi\right) \quad \text{tr}\left(\Psi^{-1} \frac{\partial}{\partial z} \Psi\right)\right] \tag{29}$$

The gradient of the tracking performance $\nabla s^n$ is then obtained as a linear combination of $\nabla c^{n,l}$ as

$$\nabla s^n = \sum_{l=1}^{L} w^l \, \nabla c^{n,l} \quad \text{for } l \in \mathbb{T}^n \tag{30}$$

following the definition of $s^n$, given in (19).

*4.2. Guidance of the Deployed Sensors*

Since the positions of the targets are changing, the performance of tracking the targets is changing too. An optimal position for tracking a group of targets at a time step will no longer be the best position in next time step. Thus, the fusion center needs to determine the next position of the sensors to chase the moving targets every time the target state is estimated. It is essentially an optimization problem similar to the discussion of the previous subsection. An intuitive way to generate commands to the sensors is to solve the optimization problem and take the solution as the position command as follows.

**Algorithm 2.1 (Guidance of sensors).**
*For* each deployed sensor $n$,
 Solve the following convex optimization problem:
  minimize $s_{Z,k+1}^n(\boldsymbol{p}_{k+1}^n)$
  subject to $\boldsymbol{p}_{k+1}^n \in P_{k+1}^n$
 Take the solution $(\boldsymbol{p}_{k+1}^n)^*$ as the position command to the sensor
**END**

The algorithm given above seems straightforward. It can be solved by the conditional gradient method addressed in (25-27). However, one can imagine that it will take time for the unmanned aircraft to reach the input position. When the targets are not stationary, the optimal position for the sensor is not stationary as well. Thus, the optimal position should be calculated at every time when new measurements are available. Moreover, executing Algorithm 2.1, i.e., solving $N$ nonlinear optimization problems, requires high computational power.

An alternative to Algorithm 2.1 is proposed as follows.

**Algorithm 2.2 (Guidance of sensors - proposed).**
*For* each deployed sensor $n$,
 Solve the following linear optimization problem:
  minimize $\nabla s_{Z,k+1}^n(\boldsymbol{p}_{k+1}^n) \cdot (\boldsymbol{p}^n - \boldsymbol{p}_{k+1}^n)$
  subject to $\boldsymbol{p}^n \in P^n$
 Take the solution $\overline{\boldsymbol{p}^n}$ as the position command to the sensor
**END**

Note that the linear optimization problem in Algorithm 2.2 is the subproblem of the optimization problem in Algorithm 2.1, revisited as

$$\begin{aligned} & \text{minimize } s_{Z,k+1}^n(\boldsymbol{p}_{k+1}^n) \\ & \text{subject to } \boldsymbol{p}_{k+1}^n \in P_{k+1}^n \end{aligned} \tag{31}$$

Each iteration of the conditional gradient method is executed at every time step for Algorithm 2.2. The sensor, which is currently at $\boldsymbol{p}_{k+1}^n$, is supposed to move toward the furthest feasible point $\overline{\boldsymbol{p}^n}$. The conditional gradient method is still valid even if the sensor couldn't reach the point $\overline{\boldsymbol{p}^n}$ at the next time step. Moving in the feasible direction, each sensor will be positioned at $\boldsymbol{p}_{k+2}^n = \boldsymbol{p}_{k+1}^n + \alpha_{k+1}(\overline{\boldsymbol{p}^n} - \boldsymbol{p}_{k+1}^n)$ where $\alpha_{k+1} \in (0,1]$. This way, the sensors always can be at feasible points which are the current best positions, requiring less computation because the algorithm is to solve a set of linear programming. Moving along the shortest path to the optimal position, by Algorithm 2.1, would take the minimum time to arrive. However, the points on the shortest path are not guaranteed feasible, so that the sensor may fail to track some targets when moving on the way to the optimal position.

The graphical illustration of the guidance algorithms is given in Fig. 2. Given a configuration of the target tracking of a sensor, the performance score can be plotted as a function of the sensor position. For simplicity, the performance score is presented as a function of x-y coordinates. At the bottom of the graph is the contour of the performance score. The solid line and dashed line denote the paths of the sensor guided by Algorithm 2.2 and Algorithm 2.1, respectively. As the dashed arrow shows, Algorithm 2.1 directs the sensor to its optimal position and the sensor is taking the shortest path to its destination. However, sensing on some points on the way may provide the higher (worse) performance and there is a possibility that the sensor is on the infeasible point, at which some of the targets are out of the sensor's coverage. On the other hand, Algorithm 2.2 directs the sensor in the feasible direction at each time step, and finally to the optimal sensor position.

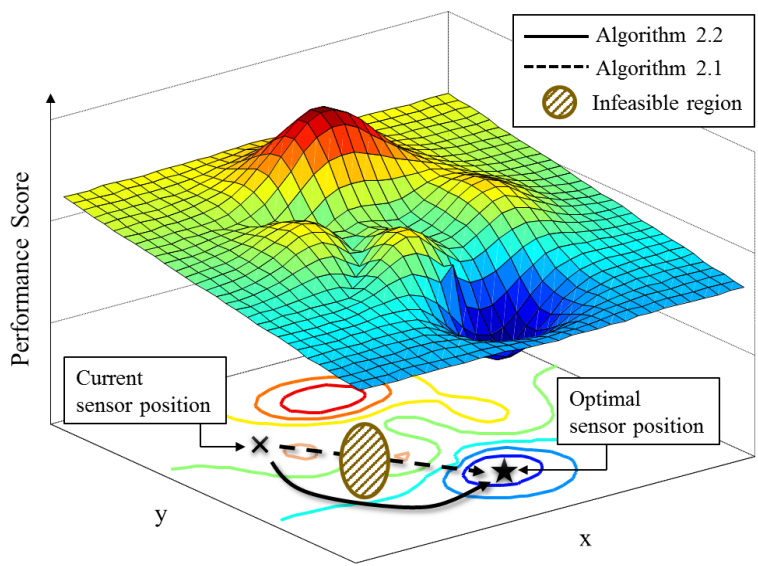

**Figure 2.** Graphical illustration of the guidance algorithms.

## 5. Simulation

We simulated the problem of tracking ground targets by sensors equipped on multiple unmanned aircraft. There are 5 unmanned aircraft and 10 targets, i.e., $N_{max} = 5$ and $L = 10$. Fig. 3 shows the targets at their initial positions. The squares denote the targets and their numbers are presented at the upper right side of the squares. The squares drawn with a thinner line denote the targets that the operator attempts not to track. The maximum number of targets a sensor can track is set $V = 4$ and the operator attempts not to track the targets 3, 4, 5, i.e., $w = [1, 1, 0, 0, 0, 1, 1, 1, 1, 1]$. The sensors are not allowed to be below $500\ m$, i.e., $H_{min} = 500\ m$. It is assumed that the new deployment is available 5 seconds after the last deployment. The time for the new deployment is determined as the time when the information score is larger than $S_1 = 4$. The initial information score is $c_0^l = 0$ for each target $l = 1, \dots, L$ since the initial covariance matrix for target state is set as $P_0^l = \mathrm{diag}([1, 1, 1, 0.25, 0.25, 0.25])$ and therefore the position part of the FIM becomes $\Psi_0^l = \mathrm{diag}([1, 1, 1])$ where $J_0^l = (P_0^l)^{-1}$. The information score will increase as the time passes if no sensor is tracking the target until the score meets $S_1$. The standard deviation of accelerometer noise is $[\sigma_x^l, \sigma_y^l, \sigma_z^l] = [0.1, 0.1, 0.1]\ m/s^2$ and the measurement is assumed to be corrupted by the noise with the standard deviation of $[\sigma_{Az}^n, \sigma_{El}^n, \sigma_R^n] = [0.007\ rad, 0.007\ rad, 0.05\ m]$.

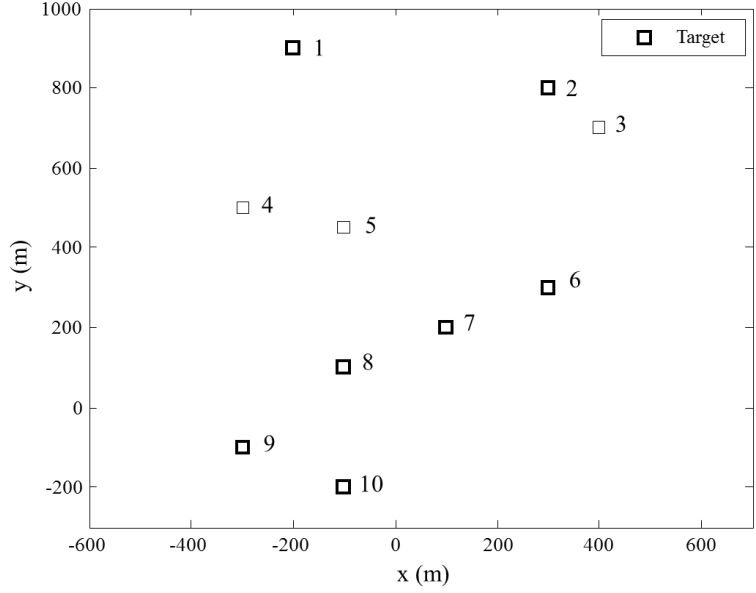

**Figure 3.** The initial configuration of the targets.

*5.1. Stationary Targets, no Guidance*

Firstly, let's see how the management algorithm works in the cases with stationary targets. In the first case, the guidance algorithm (Algorithm 2.2) will not be executed after the configuration of the new sensor is determined by Algorithm 1. Thus, the deployed sensors will remain stationary at their optimal position. Fig. 4 shows the time history of information about each target. Note that lower value of the information score means better tracking performance. The dashed lines denote the threshold set by the operator. If the information score meets the threshold, the fusion center deploys the new sensor. The fusion sensor predicted that the information of the target 1 will be insufficient at $t = 7.85\ s$, and consequently the first sensor was deployed at $\boldsymbol{p}^{n=1} = [-149, 358, 1105]^T\ m$ to track the targets $\{1, 7, 8, 9\}$ from that time on. It can be observed that the information score of the targets being tracked has decreased after the deployment. That is, the Algorithm 1 determined that the configuration provides the best target tracking performance score. The next deployment was occurred at $t = 12.85\ s$. As a result, the final target groups $\mathbb{T}_f$ for all sensors can be expressed as an $N \times L$ matrix as

$$\mathbb{T}_f = \begin{bmatrix} 1 & 0 & 0 & 0 & 0 & 0 & 1 & 1 & 1 & 0 \\ 0 & 1 & 0 & 0 & 0 & 1 & 1 & 0 & 0 & 1 \end{bmatrix} \tag{32}$$

where 1 in *n*th row and *l*th column in $\mathbb{T}_f$ means that the target *l* is being tracked by the sensor *n*. The final configuration is shown in Fig. 5.

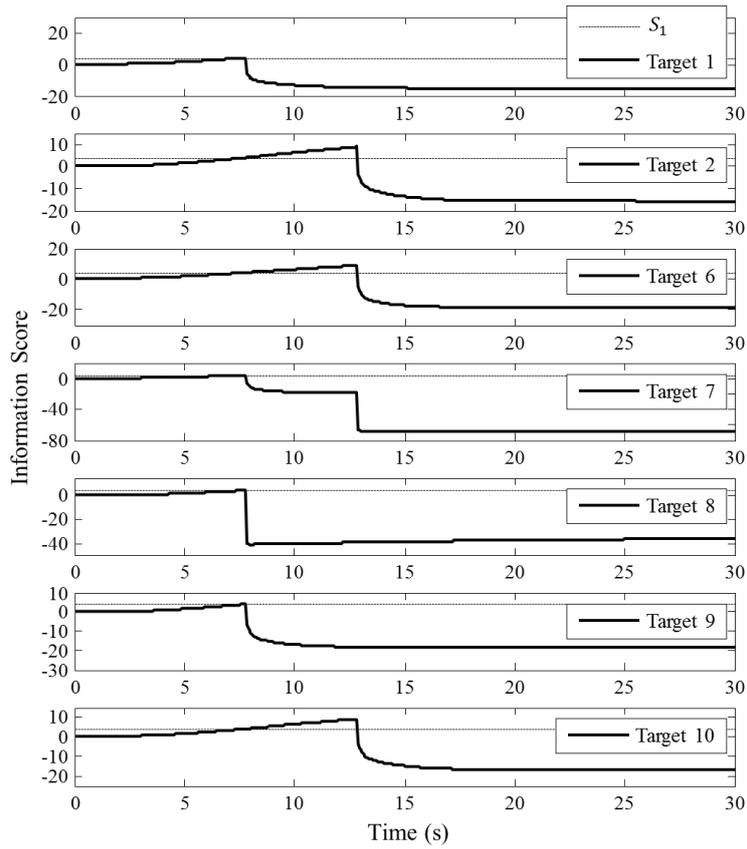

**Figure 4.** The time history of the information score of each target when the targets are stationary and the sensors are deployed at their optimal positions.

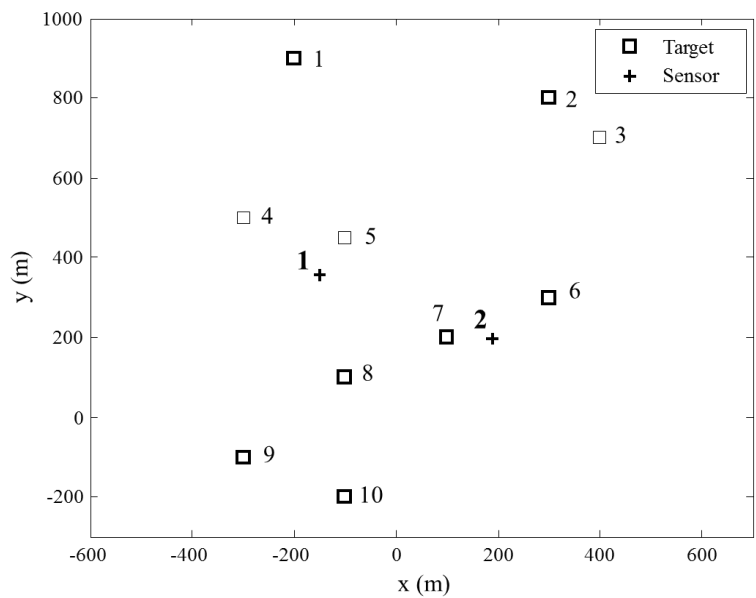

**Figure 5.** The final locations of the targets and the deployed sensors for case 5.1.

*5.2. Stationary Targets, with Guidance*

As the second case with the stationary targets, the sensors are deployed at the places that are not optimal for the tracking performance and they are supposed to track the targets assigned as

$$\mathbb{T}_0 = \begin{bmatrix} 1 & 0 & 0 & 0 & 0 & 1 & 1 & 1 & 0 & 0 \\ 0 & 1 & 0 & 0 & 0 & 0 & 1 & 1 & 1 & 0 \\ 0 & 1 & 0 & 0 & 0 & 0 & 0 & 0 & 1 & 1 \end{bmatrix} \tag{33}$$

Throughout the simulation, the sensor movement is assumed as the first-order lag model whose speed is limited by 3 m/s for each axis. By executing the guidance algorithm (Algorithm 2.2), we observe the behavior of the sensors as Fig. 6. The sensors were directed to the optimal position and the performance score has the lower value as each sensor moves toward the optimal position as depicted in Fig. 7.

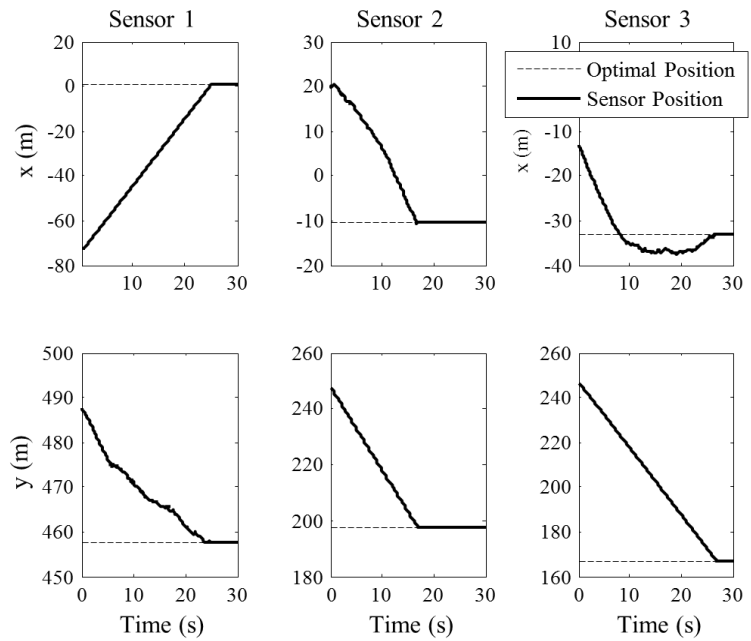

**Figure 6.** The time history of the sensor positions drawn with their optimal positions when each sensor is deployed at a random position and guided by Algorithm 2.2.

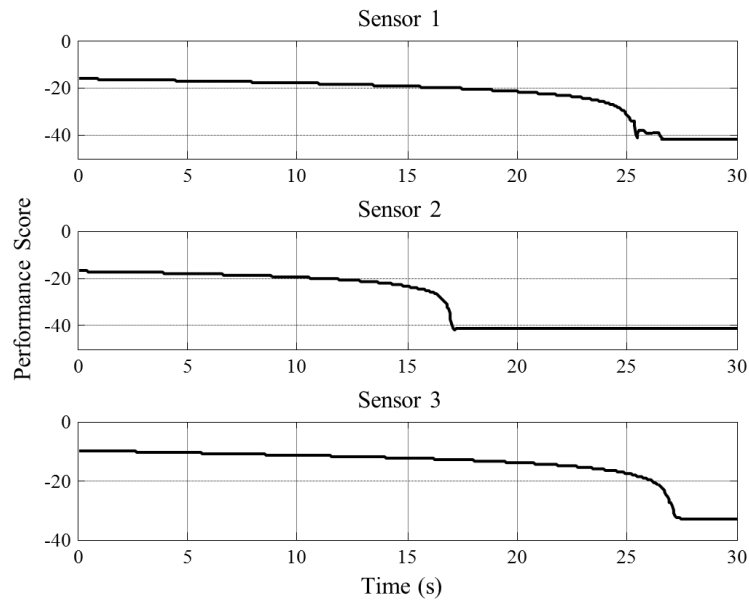

**Figure 7.** The time history of the performance score of each sensor guided by Algorithm 2.2 to track the stationary targets.

*5.3. General Case*

In this case, each target is moving independently with the speed of $0 - 3$ m/s. Initially, three sensors are assigned to track the targets as

$$\mathbb{T}_0 = \begin{bmatrix} 1 & 0 & 0 & 0 & 0 & 0 & 0 & 0 & 0 & 0 \\ 0 & 1 & 0 & 0 & 0 & 0 & 1 & 0 & 0 & 0 \\ 0 & 0 & 0 & 0 & 0 & 0 & 1 & 1 & 1 & 0 \end{bmatrix} \quad (34)$$

The fusion center executes Algorithm 1 (Configuration of new deployment) and Algorithm 2.2 (Guidance of sensors) as depicted in Fig. 1. The resultant information score of the targets are plotted against time as Fig. 8. It is observed that, at $t = 7.85\ s$, the information of the target 10 was considered insufficient and the targets $\{8, 9, 10\}$ started being tracked by the newly deployed sensor 4. After the deployment, there was no target whose information score is larger than the threshold $S_1$.

The time history of performance score of each sensor is plotted in Fig. 9, compared with those in Algorithm 2.1 and with no guidance command. In the case with no guidance, the sensors are being stationary at their initial position. We observe that the performance scores of the sensors are increasing as the time passes when no guidance algorithm is executed while the sensors guided by Algorithm 2.1 or Algorithm 2.2 provide better (lower) performance scores.

Note that Algorithm 2.2 gives better performance than Algorithm 2.1 which solves the optimization problem for the performance score of a sensor and takes the optimal solution as the guidance command. When the targets are moving, the optimal position to track the targets is changing and the sensors may not make it to the optimal position at next time steps. Instead, the sensors guided by Algorithm 2.2 move in the furthest feasible direction and achieve the current best performance score. This tendency is same for the most cases in the simulations except the cases with stationary targets. When the targets are stationary, Algorithm 2.1 shows the better performance since Algorithm 2.1 guides each sensors directly to the (fixed) optimal position and the sensor takes the shortest path to its destination.

Furthermore, Algorithm 2.1 requires much more computation time than Algorithm 2.2. It is obvious since Algorithm 2.1 is supposed to iteratively solve its subproblem (31) which is same to Algorithm 2.2 itself. Thus, the required computation time is proportional to the number of iterations in Algorithm 2.1. In the simulation, the average computation time of executing Algorithm 2.1 was 74 times longer than that of executing Algorithm 2.2 that implies that the average number of iterations in Algorithm 2.1 was about 74.

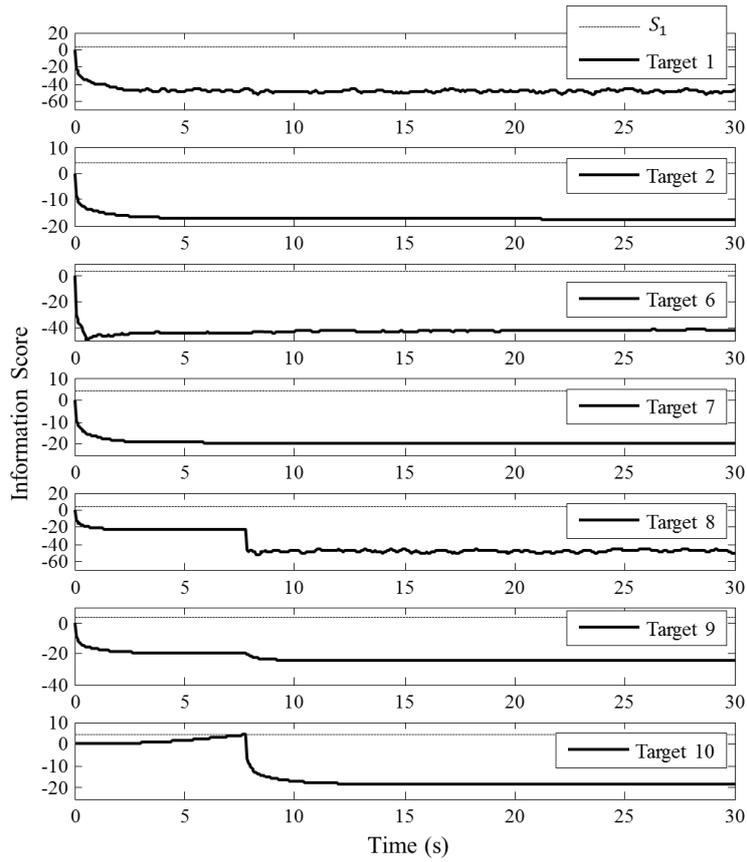

**Figure 8.** The time history of the information score of each target when the targets are moving and the sensors are deployed at their optimal positions.

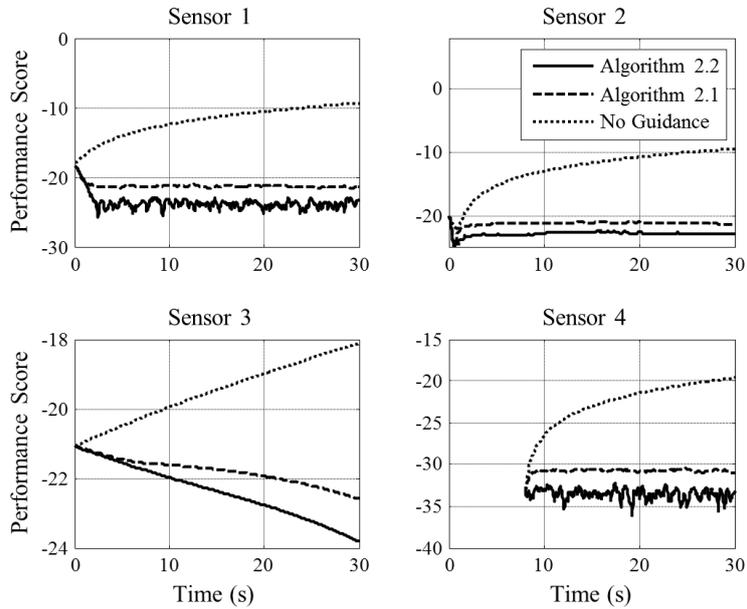

**Figure 9.** The performance scores of the sensors resulted in being guided by Algorithm 2.2 (solid line), Algorithm 2.1 (dashed line) are plotted against time, with the results with no guidance command (dotted line).

## 6. Conclusion

This paper proposed a framework for airborne multisensor management for target tracking. The performance metric was newly defined and its convexity was proved. The convex optimization was used in determining the configuration of the new sensor deployment as well as in the guidance of the deployed sensors to improve their target tracking performance.

In providing optimal position to each sensor, an intuitive way that directs the sensor to its optimal position by solving the nonlinear optimization problem did not perform well as the targets are moving. Instead, we proposed an algorithm that provides the negative gradient of the cost function by solving the subproblem of the nonlinear optimization problem. The simulation result showed that the proposed algorithm obtained better performance score requiring much less computation time.

In conclusion, the proposed framework addressed in this paper is a powerful tool that allows the efficient management of the unmanned aircraft as sensors.